# Twinning Relationships in Coexisting Cubic and Tetragonal Phases of Ferroelectrics

Semën Gorfman[1], Ido Biran[1], Nan Zhang[2], Zuo-Guang Ye[3]

[1] Department of Materials Science and Engineering, Tel Aviv University, Tel Aviv, Israel

[2] Electronic materials research laboratory, Xi'an Jiaotong University, Xi'an, China

[3] Department of Chemistry and 4D LABS, Simon Fraser University, Burnaby, British Columbia, V5A 1S6, Canada

**Abstract**

Many ferroelectric materials undergo thermally driven transitions between cubic and tetragonal phases. Despite being well-known for decades, such transitions still pose unresolved questions, particularly about the role of domains in shaping the transition pathway. It is understood that tetragonal domain assemblages are crucial for mismatch-free coexistence of both tetragonal and cubic phases, as described by the Weschler—Lieberman—Read (WLR) crystallographic theory. However, direct experimental characterization of domain patterns during the phase transition remains challenging. Here, we enhance the single-crystal X-ray diffraction to investigate domain microstructures during cubic-tetragonal phase transition, leveraging the bulk-penetrating and non-destructive character of this method. We extend the WLR model by the analytical expressions for the orientation relationships between the real and reciprocal bases of the cubic phase and tetragonal domains. We systematically catalogue the Bragg peak separations for 24 possible coexistence variants and validate these predictions using 3D reciprocal space maps from a piezo-/ferroelectric PMN-35PT crystal at phase coexistence temperatures. Our results provide a framework for interpreting domain configurations during the phase transition and offer insights applicable to other systems involving orientation relationships in coexisting phases.

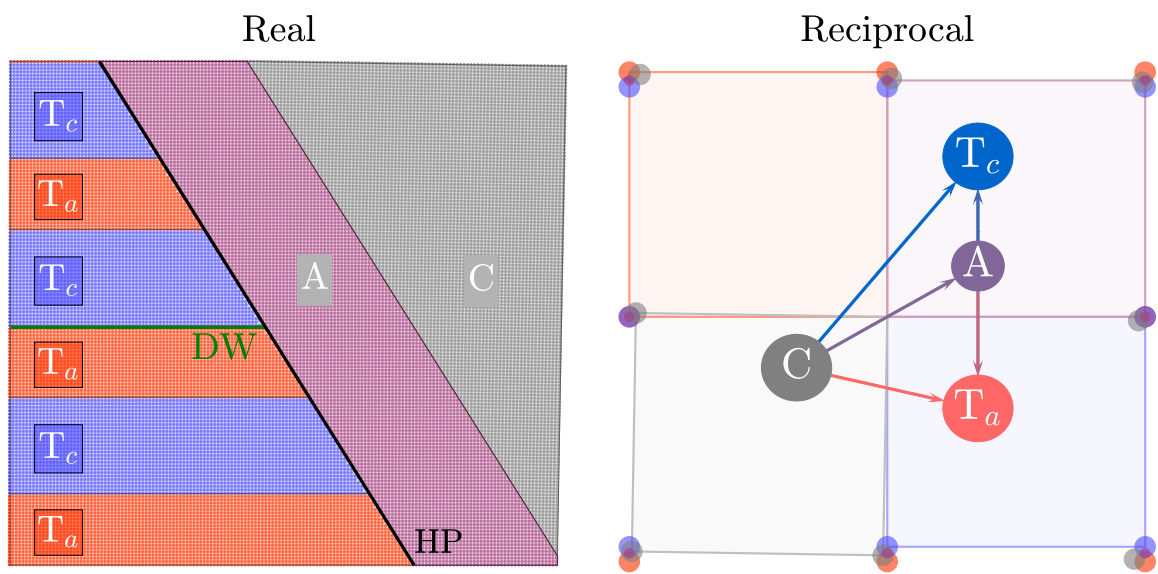

## 1. Introduction

The transition from the cubic (C) paraelectric to the tetragonal (T) ferroelectric phase is one of the most prevalent occurrences in perovskite-based materials. It is exemplified by BaTiO$_3$ at approximately 125°C [1,2], but also observed in many other ferroelectric materials, such as KTa$_{1-x}$Nb$_x$O$_3$ [3], PbZr$_{1-x}$Ti$_x$O$_3$ [4], Na$_{0.5}$Bi$_{0.5}$TiO$_3$ [5] and Pb(Mg$_{1/3}$Nb$_{2/3}$)O$_3$ - PbTiO$_3$ [6]. Like any other symmetry-lowering phase transition [7–10], it leads to the emergence of domain microstructures, which in turn underpin such physical properties as ferroelasticity [11], piezoelectricity [12] and shape-memory effect [13]. Such CT phase transition involves the change in space group symmetry in general, but more specifically transforms the conventional C-lattice parameters ($a_C$ $a_C$ $a_C$ 90 90 90) to the T- ones ($a_T$ $a_T$ $c_T$ 90 90 90) [14]. For the case when the transition is of first-order type, both phases may coexist in a certain temperature interval. Despite being recognized for six decades, the multi-length-scale nature of such transitions remains poorly understood and the functionalization potential of phase coexistence remains unexplored.

A significant challenge in describing the CT phase transition lies in the lattice mismatch between potentially coexisting C and T phases. It is impossible to find a lattice plane that could serve as a mismatch-free interface, unless $a_T = a_C$. This means that the CT coexistence involves a lateral strain at the phase separation interface. One proposed model to mitigate this strain introduces the so-called "adaptive state" (AS) in the tetragonal [15–18] or any other lower symmetry (e.g. rhombohedral or orthorhombic) phases. AS is a periodic assemblage of alternating T-domains that can be averaged into a pseudo-uniform phase of monoclinic symmetry that, in turn, allows a mismatch-free connection to the C-phase. This concept is also known as Wechsler—Lieberman—Read (WLR) crystallographic theory (e.g. [19]) which was introduced to describe similar phase transitions in martensitic materials [13,20–22], the paraelastic-ferroelastic interfaces in ferroelastics [23], the properties of domain walls if perovskites [24]. Both AS and WLR models assume miniaturization of domains which justifies their macroscopic averaging. Yet, the quantitative meaning of the term "miniaturization" is either absent or uncertain.

The lack of suitable experimental techniques means that the observations of AS and related phenomena in three-dimensional bulk are rare. The majority of experimental techniques are sensitive to the surface only (e.g. PFM), applicable to very small samples (e.g. TEM) or provide too limited crystallographic information (as e.g. polarized light / birefringence optical microscopy [25],[26] or confocal Raman spectroscopy [27]). Single crystal X-ray diffraction stands out by providing non-destructive and in-bulk exploration of the multi-length scale phenomena. Specifically, three-dimensional reciprocal space mapping (3D-RSM) gives some more substantial information about domain microstructures [28–30] and their alternation during temperature-driven phase transitions or under an external electric field. These capabilities are further enhanced by the synchrotron-based Dark-Field-X-Ray-Microscopy methods [31–33]. Nevertheless, the complexity of the domains and the corresponding X-ray diffraction patterns has set up a persisting challenge for the analysis of domain structures and twin relations.

This paper introduces a mathematical framework that enables the investigation of CT coexisting mechanisms using high-resolution X-ray diffraction 3D reciprocal space mapping. This builds up on the previous works [34–36] towards enabling investigation of coherent twin relationship in ferroelectrics. We introduce the model of CT coexistence via the periodic assemblage of T-domains as in the AS, but without necessarily domains miniaturization. Based on this model, we catalog all the variants of such coexistence and predict the splitting of Bragg peaks in 3D reciprocal space. We compare the predictions with the experimental results from high-resolution reciprocal space mapping of PMN-35PT and show that this comparison supports the formation of AS-like state without domains miniaturization.

## 2. The organization of the paper

All the relevant notations and crystallographic relationships are summarized in Appendices A1 and A2. The paper begins by recapitulating the most important AS features, including the underlying twinning of T-domains, the description of their macroscopic averaging, and the calculation of the corresponding average lattice parameters. Then, we consider possible mismatch-free ASC connections and derive the relevant orientation relationship between the direct and reciprocal crystallographic basis vectors. We then catalog all the possible ASC cases and predict the reciprocal space separation between the Bragg peaks for each of them. Finally, we validate the introduced model using the experimental results from 3D reciprocal space mapping of a multi-domain PMN-35PT crystal.

## 3. Tetragonal domains and permissible domain walls between them

The CT phase transition leads to the change of the C-lattice parameter $a_C a_C a_C$ 90 90 90 into T-lattice parameters $a_T a_T c_T$ 90 90 90 (here and everywhere in the paper, the pseudocubic cell setting is used). The corresponding tetragonal distortion is quantified by the dimensionless quantities $x_a, x_c, x_T,$ and $\tau$ defined in (A1.1 – A1.3). Three ferroelastic tetragonal domain variants are referred to as $T_a$, $T_b$, and $T_c$ as elaborated in (A1.5). These variants can pair with one another along six permissible domain walls (PDW) [34,37,38], where the word "permissible" indicates the absence of any lattice mismatch along the connection interface. Table 1 lists these pairings: it includes T-domains involved in the pair, the Miller indices of PDW between them and the orientation relationship between the pseudocubic basis vectors. This relationship is expressed by the corresponding transformation matrix $[\Delta S]_{mn} = [S]_{mn} - [I]$ in (A1.6) and (A1.7).

**Table 1**. Permissible Domain Walls (PDWs) describing the mismatch-free pairings of T-domains. Columns 1 and 2 define the domain pairs. Each pair of domains may have two PDWs. Columns 3 and 5 contain the Miller indices of the PDW. Columns 4 and 6 define the orientation relationship $[\Delta S]_{mn}$ between the pseudocubic basis vectors according to (A1.6) and (A1.7). The tetragonal distortion parameter $\tau = \frac{c_T^2 - a_T^2}{c_T^2 + a_T^2}$.

| $m$ | $n$ | Plane 1 | $[\Delta S]_{mn}$ | Plane 2 | $[\Delta S]_{mn}$ |
|---|---|---|---|---|---|
| $T_a$ | $T_b$ | $(1\bar{1}0)$ | $\begin{pmatrix} \bar{\tau} & \tau & 0 \\ \bar{\tau} & \tau & 0 \\ 0 & 0 & 0 \end{pmatrix}$ | $(110)$ | $\begin{pmatrix} \bar{\tau} & \bar{\tau} & 0 \\ \tau & \tau & 0 \\ 0 & 0 & 0 \end{pmatrix}$ |
| $T_a$ | $T_c$ | $(10\bar{1})$ | $\begin{pmatrix} \bar{\tau} & 0 & \tau \\ 0 & 0 & 0 \\ \bar{\tau} & 0 & \tau \end{pmatrix}$ | $(101)$ | $\begin{pmatrix} \bar{\tau} & 0 & \bar{\tau} \\ 0 & 0 & 0 \\ \tau & 0 & \tau \end{pmatrix}$ |
| $T_b$ | $T_c$ | $(01\bar{1})$ | $\begin{pmatrix} 0 & 0 & 0 \\ 0 & \bar{\tau} & \tau \\ 0 & \bar{\tau} & \tau \end{pmatrix}$ | $(011)$ | $\begin{pmatrix} 0 & 0 & 0 \\ 0 & \bar{\tau} & \bar{\tau} \\ 0 & \tau & \tau \end{pmatrix}$ |

## 4. Adaptive state and macroscopic averaging of T-domains

Figure 1 illustrates the adaptive state (AS) variant made of the $T_a(110)T_b$ pair, connected along the (110) PDW with the different volume ratios of each domain. Figure 2 emphasizes the concept of their macroscopic averaging, showing the multi-domain state (a), the average state (b), and the overlay of both (c).

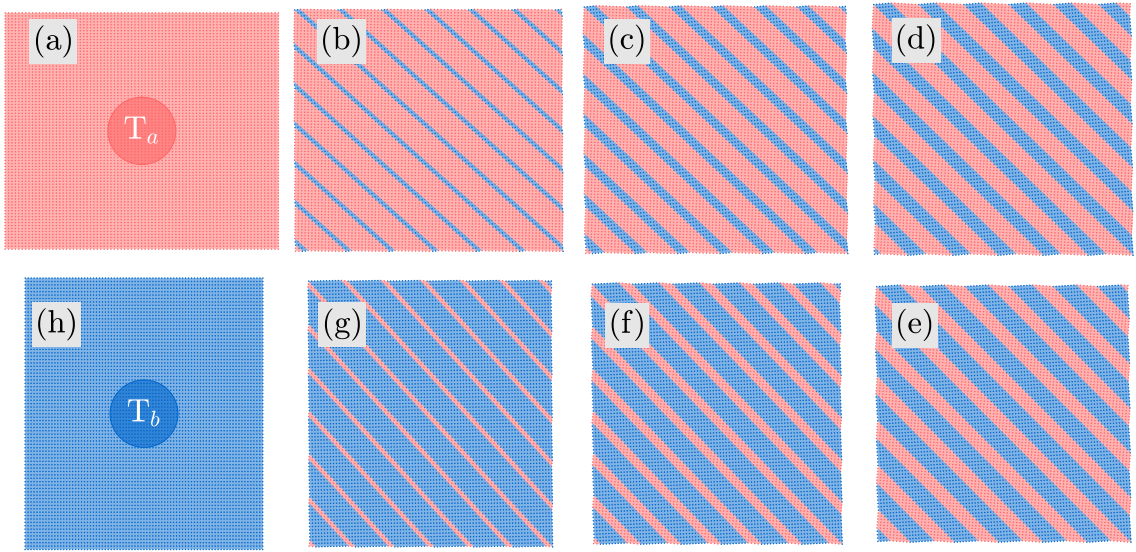

**Figure 1.** Schematic two-dimensional illustration of the adaptive state (AS) concept. The AS is a domain configuration where two domain variants (e.g., $T_a$ and $T_b$) are connected along a specific PDW, here identified as (110). The domains form a periodic pattern with a fixed domain volume ratio. Panels (a) and (h) represent pure $T_a$ and $T_b$ domains state. Panels (b) to (g) illustrate mixed states, with the volume fraction of $T_b$ progressively increasing from 0 (a) to 1 (b).

We will use the parameter $\omega$ to describe the domain volume ratio, e.g. domain $T_b$ in the $T_a(110)T_b$ AS variant. Figure 3 zooms in on the $T_a(110)T_b$ connection with the $T_a$ / $T_b$ domain volume ratios of $1 - \omega$ and $\omega$, respectively.

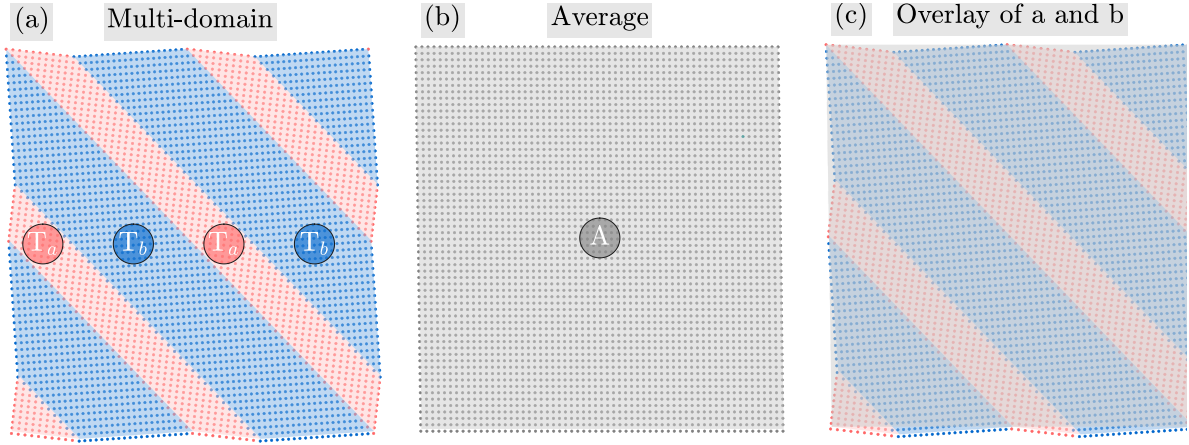

**Figure 2.** Two-dimensional schematic illustration of the adaptive state (AS) concept, emphasizing the macroscopic averaging. Panel (a) depicts the original multidomain configuration, which is averaged to yield a macroscopically uniform state shown in Panel (b). Panel (c) overlays the multidomain state and its macroscopic average state for comparison.

Both "AS" and "WLR" models assume that the macroscopic averaging of these domains creates a macroscopically uniform state with the average basis vectors $\boldsymbol{a}_{iA} = (1-\omega)\boldsymbol{a}_{im} + \omega\boldsymbol{a}_{in}$, as illustrated in Figure 3. The transformation of crystallographic basis vectors $\boldsymbol{a}_{im} \to \boldsymbol{a}_{iA}$ is described by the matrix $[\Delta S]_{mA}$ (see Appendix A for the formal definitions of these transformation matrices).

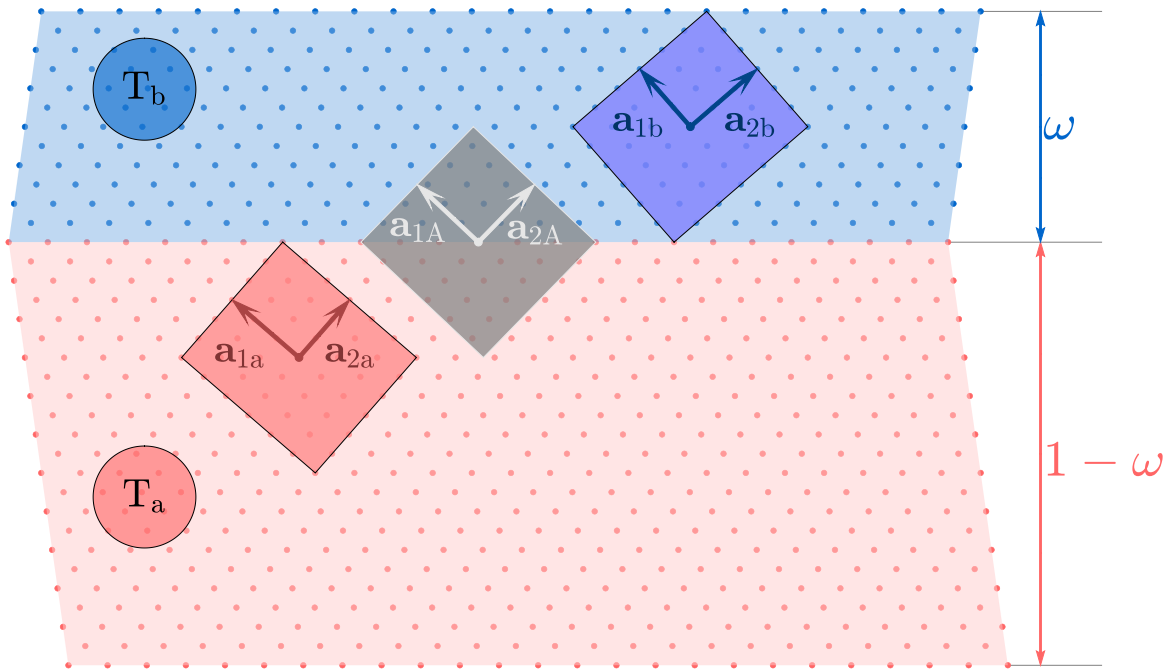

**Figure 3**. A schematic illustration of two tetragonal domain variants, $T_a$ and $T_b$, connected along the (110) permissible domain wall. The volumetric ratios of these domains are $1 - \omega$ and $\omega$, respectively, as indicated in the figure. The figure includes the [uv0]-section of the pseudocubic unit cells, the corresponding basis vectors for each domain, and the "average" unit cell. These unit cells are enlarged for clarity.

$$[\Delta S]_{mA} = \omega [\Delta S]_{mn} . \tag{1}$$

The average lattice parameters of the AS-state can be calculated using the corresponding matrix of dot products $G_{ij,A} = \boldsymbol{a}_{iA} \cdot \boldsymbol{a}_{jA}$. According to (A1.8), $[G]_A = [S]_{mA}^T [G]_m [S]_{mA}$, so that using $[S]_{mA} = [I] + [\Delta S]_{mA}$ we get:

$$[G]_A = [G]_m + \omega [\Delta S]_{mn}^T [G]_m + \omega [G]_m [\Delta S]_{mn} + \omega^2 [\Delta S]_{mn}^T [G]_m [\Delta S]_{mn} . \tag{2}$$

Let us consider the $T_a(110)T_b$ AS variant with $m = $ a and $n = $ b as presented in the second row of Table 1 and in Figures 1, 2 and 3. According to (A1.5) and Table 1:

$$[G]_m = [G]_a = \begin{pmatrix} c_T^2 & 0 & 0 \\ 0 & a_T^2 & 0 \\ 0 & 0 & a_T^2 \end{pmatrix}, [\Delta S]_{mn} = \begin{pmatrix} \bar{\tau} & \bar{\tau} & 0 \\ \tau & \tau & 0 \\ 0 & 0 & 0 \end{pmatrix} . \tag{3}$$

Using (2) and (3) we can express $[G]_A$ as the Taylor series expansion with respect to the small tetragonal distortion parameter $\tau$:

$$[G]_A = a_T^2 [I] + \tau [G]_{A1} + \tau^2 [G]_{A2} + O(\tau^3) . \tag{4}$$

The first term $a_T^2 [I]$ does not depend on $\tau$, while the second term, $\tau [G]_{A1}$, does, and $[G]_{A1}$ can be expressed as:

$$[G]_{A1} = (c_T^2 + a_T^2) \begin{pmatrix} 1 & 0 & 0 \\ 0 & 0 & 0 \\ 0 & 0 & 0 \end{pmatrix} + 2a_T^2 \omega \begin{pmatrix} \bar{1} & 0 & 0 \\ 0 & 1 & 0 \\ 0 & 0 & 0 \end{pmatrix} . \tag{5}$$

The diagonal form of $a_T^2 [I] + \tau [G]_{A1}$ suggests that the corresponding "AS-lattice" has an orthorhombic symmetry with all different pseudocubic axes lengths and all the angles equal to 90°. Adding the second-order term $\tau^2 [G]_{A2}$ lowers this symmetry to the monoclinic $M_C$ type with the angles between $\boldsymbol{a}_{1A}$ and $\boldsymbol{a}_{2A}$ different from 90°. For the illustrative purposes and to enable analytical description of ASC coexistence variants, we will neglect all the terms containing the second or higher order of $x_a, x_c,$ and $\tau$. Starting from the next paragraph this will be expressed by the sign " $\approx$ ", instead of " $=$ ".

It is worth noting that this reduction in average symmetry to monoclinic (or orthorhombic) has been widely debated in literature, particularly in the context of monoclinic ferroelectric phases — first reported by [39]. A central question is whether these monoclinic ferroelectric phases represent true long-range ordered states [40,41] or if they solely arise as a result of macroscopic averaging of domains. A detailed discussion of this issue lies beyond the scope of the present work, because the monoclinic ferroelectric phases are introduced in a significantly broader context. Here, we employ the concept of a low symmetry phase solely as a tool to describe the mismatch-free CT coexistence only, as further elaborated in the following paragraph.

## 5. "Habit Plane" between adaptive state and cubic phase.

The Habit Plane (HP) is a mismatch-free planar interface between the adaptive state (AS) and the cubic (C) phase. Here the term "mismatch-free" stands for exact matching of two-dimensional lattice parameters within the interface plane. For AS these lattice parameters are defined according to (4) and (5). Appendix B describes the procedure for the calculation of the "Miller" indices of such interface, regardless of if this interface connects pure or adaptive states. The mismatch-free interface between AS state and C phase (ASC - connection) is commonly referred to as Habit Plane (HP). The existence of HP for an ASC connection is subject to meeting the necessary condition [34]:

$$\det[\Delta G]_{AC} = 0, \tag{6}$$

where $[\Delta G]_{AC} = [G]_A - [G]_C$, with $[G]_C = a_C^2[I]$. According to (4) and for the $T_a(110)T_b$ AS variant, $[\Delta G]_{AC}$ has three non-zero elements:

$$\begin{aligned}\Delta G_{11,AC} &\approx (c_T^2 - a_C^2) - 2a_T^2\omega\tau, \\ \Delta G_{22,AC} &\approx (a_T^2 - a_C^2) + 2a_T^2\omega\tau, \\ \Delta G_{33,AC} &= (a_T^2 - a_C^2).\end{aligned} \tag{7}$$

If $a_T \neq a_C$ then (6) is fulfilled if either $\Delta G_{11,AC} = 0$ or $\Delta G_{22,AC} = 0$. These conditions can be satisfied by tuning $\omega$ to $\omega_1$ and $\omega_2$, respectively:

$$\omega_1 = \frac{c_T^2 - a_C^2}{2a_T^2\tau} \approx \frac{x_c}{x_T}, \tag{8}$$

or

$$\omega_2 = \frac{a_C^2 - a_T^2}{2a_T^2\tau} \approx \frac{\bar{x}_a}{x_T}. \tag{9}$$

Accordingly, two variants of the mismatch-free ASC connections exist. Here, we will show the derivations for $\omega = \omega_1$ while the case $\omega = \omega_2$ can be processed analogously. For $\omega = \omega_1$ we get

$$[\Delta G]_{AC} \approx 2a_C^2 \begin{pmatrix} 0 & 0 & 0 \\ 0 & x_a + x_c & 0 \\ 0 & 0 & x_a \end{pmatrix}. \tag{10}$$

The volume of the AS "average unit cell" (needed later) can be calculated as $V_A^2 = \det[G]_A$:

$$V_A \approx a_C^3(1 + 2x_a + x_c). \tag{11}$$

The next step towards the calculation of the "Miller indices" of HP (Appendix B) is to find the non-zero eigenvalues of $[\Delta G]_{AC}$:

$$\lambda_1 = 2a_C^2 x_a,$$
$$\lambda_3 = 2a_C^2 (x_a + x_c). \tag{12}$$

And the corresponding orthogonal matrix of eigenvectors (arranged column-wise):

$$[V] = \begin{pmatrix} 0 & 1 & 0 \\ 0 & 0 & 1 \\ 1 & 0 & 0 \end{pmatrix}. \tag{13}$$

According to Appendix B, the existence of HP is subject to the condition $\lambda_1 \lambda_3 < 0$. Assuming that $x_a < 0$ ($a_T < a_C$), we can express the condition for the existence of HP as:

$$x_c > |x_a|. \tag{14}$$

The "Miller indices" of HP can be derived according to the equation (B1.7):

$$\begin{aligned} p_{i,1} &= \Lambda V_{i1} - V_{i3} \\ p_{i,2} &= \Lambda V_{i1} + V_{i3} \end{aligned}, \tag{15}$$

where:

$$\Lambda = \left(\frac{|\lambda_1|}{|\lambda_3|}\right)^{\frac{1}{2}}. \tag{16}$$

For the specific connection considered here we can insert (12) into (16) and obtain $\Lambda = M$, with

$$M = \left(\frac{\bar{x}_a}{x_a + x_c}\right)^{\frac{1}{2}}. \tag{17}$$

Using (13) and (15), we get

$$[p]_1 = \begin{pmatrix} 0 \\ \bar{1} \\ M \end{pmatrix}$$
$$[p]_2 = \begin{pmatrix} 0 \\ 1 \\ M \end{pmatrix}. \tag{18}$$

Table 2 highlights several favorable cases which set the HP parallel to a lattice plane with rational Miller indices. For example, $x_c = -2x_a$ means the presence of $(011)$- or $(0\bar{1}1)$ − HPs. Approaching $x_a = 0$ (e.g., $a_T = a_C$) means the presence of the $(011)$ HP.

**Table 2**. Favorable T-distortion parameters setting the mismatch-free ASC interface parallel to the lattice planes with small Miller indices.

| Conditions | $p_1$ | $p_2$ |
| --- | --- | --- |
| $2x_a + x_c = 0$ | $(0\bar{1}1)$ | $(011)$ |
| $x_a + x_c = 0$ | $(001)$ | $(001)$ |
| $5x_a + x_c = 0$ | $(0\bar{2}1)$ | $(021)$ |

$$10x_a + x_c = 0 \qquad (0\bar{3}1) \qquad (031)$$

$$x_a = 0 \qquad (010) \qquad (010)$$

---

## 6. The orientation relationship between the C and AS crystallographic basis vectors

This paragraph demonstrates the derivation of the transformation matrix $[\Delta S]_{CA}$ for the basis vectors $\boldsymbol{a}_{iC} \to \boldsymbol{a}_{iA}$. Knowing such orientation relationships is useful for various crystallographic calculations, specifically to obtain the separation between the Bragg peaks diffracted from coexisting domains and phases. According to [34] and Appendix B, $[\Delta S]_{CA}$ is calculated using the equation:

$$[\Delta S]_{CA} = [V][Z]\begin{pmatrix} 0 & 0 & y_1 \\ 0 & 0 & y_2 \\ 0 & 0 & y_3 - 1 \end{pmatrix}[Z]^{-1}[V]^T, \qquad (19)$$

where, $y_3$ is the ratio of the AS and C unit cell volumes according to (11):

$$y_3 \approx 1 + 2x_a + x_c, \qquad (20)$$

and

$$[Z] = \begin{pmatrix} 1 & 0 & 0 \\ 0 & 1 & 0 \\ Z_{31} & 0 & M \end{pmatrix}, \qquad (21)$$

where $Z_{31} = +M$ and $-M$ for $\boldsymbol{p}_1$ and $\boldsymbol{p}_2$, respectively. The coefficients $y_1$ and $y_2$ can be calculated from the system of equations derived in [34]:

$$\begin{pmatrix} G'_{11,C} & G'_{12,C} \\ G'_{21,C} & G'_{22,C} \end{pmatrix}\begin{pmatrix} y_1 \\ y_2 \end{pmatrix} = \begin{pmatrix} G'_{13,A} - G'_{13,C}y_3 \\ G'_{23,A} - G'_{23,C}y_3 \end{pmatrix}, \qquad (22)$$

where

$$[G']_{C,A} = [Z]^T[V]^T [G]_{C,A} [V][Z]. \qquad (23)$$

Appendix B shows that for the considered ASC variant, (22) leads to

$$\begin{aligned} y_1 &\approx \mp x_a \\ y_2 &= 0 \end{aligned}, \qquad (24)$$

where $y_1 = -x_a$ and $+x_a$ for $\boldsymbol{p}_1$ and $\boldsymbol{p}_2$, respectively. Using (19) we get for $\boldsymbol{p}_1$ ($Z_{31} = +M, y_1 = -x_a$):

$$[\Delta S]_{CA} \approx (x_a + x_c)\begin{pmatrix} 0 & 0 & 0 \\ 0 & 1 & \overline{M} \\ 0 & M & \overline{M^2} \end{pmatrix}. \qquad (25)$$

Similarly, for $\boldsymbol{p}_2$ ($Z_{31} = -M, y_1 = +x_a$):

$$[\Delta S]_{CA} \approx (x_a + x_c) \begin{pmatrix} 0 & 0 & 0 \\ 0 & 1 & M \\ 0 & \bar{M} & M^2 \end{pmatrix}. \tag{26}$$

## 7. The orientation relationship between C and T basis vectors

We will now demonstrate the derivation of the similar matrices $[\Delta S]_{Cm}$ and $[\Delta S]_{Cn}$ for the transformations $\boldsymbol{a}_{iC} \to \boldsymbol{a}_{im}$ and $\boldsymbol{a}_{iC} \to \boldsymbol{a}_{in}$. Using equation (A1.6), we will directly get:

$$[\Delta S]_{Cm} = [\Delta S]_{CA} + [\Delta S]_{Am}, \tag{27}$$

where $[\Delta S]_{Am} = [S]_{mA}^{-1} - [I]$. Using the equation (1) $[S]_{mA} - [I] = \omega[\Delta S]_{mn}$ and the fact that $[\Delta S]_{mn}^2 = 0$ holds (according to Table 1), we can obtain:

$$[\Delta S]_{Cm} = [\Delta S]_{CA} - \omega[\Delta S]_{mn}. \tag{28}$$

For the domain $n$, $[\Delta S]_{Cn} = [\Delta S]_{Cm} + [\Delta S]_{mn}$, so that:

$$[\Delta S]_{Cn} = [\Delta S]_{CA} + (1 - \omega)[\Delta S]_{mn}. \tag{29}$$

For the $T_a(110)T_b$ variant, we use $[\Delta S]_{mn} = \tau \begin{pmatrix} \bar{1} & \bar{1} & 0 \\ 1 & 1 & 0 \\ 0 & 0 & 0 \end{pmatrix}$ and consider that $\tau \approx x_T$. For the case of $\omega = \omega_1 = \frac{x_c}{x_T}$ and $1 - \omega = \frac{\bar{x}_a}{x_T}$, we get:

$$[\Delta S]_{Am} \approx x_c \begin{pmatrix} 1 & 1 & 0 \\ \bar{1} & \bar{1} & 0 \\ 0 & 0 & 0 \end{pmatrix}, \tag{30}$$

and

$$[\Delta S]_{An} \approx x_a \begin{pmatrix} 1 & 1 & 0 \\ \bar{1} & \bar{1} & 0 \\ 0 & 0 & 0 \end{pmatrix}. \tag{31}$$

## 8. Separation between the Bragg peaks diffracted from the ASC assembly

Let us consider the reciprocal lattices corresponding to the specific ASC variant comprising four differently oriented components: two corresponding to the T-, one to their average (A) and one to the C-phase, as schematically demonstrated in Figure 4a. The original definition of AS assumes the miniaturization of the constituting T-domains, which requires accounting for the interference between the X-ray (electron) beams diffracted from them [42,43]. However, such a case may only be relevant at the phase transition. Here, we introduce a model that assumes that the T-domains are large enough to diffract independently (incoherently) while still maintaining the orientation relationships imposed by the ASC-coexistence. Consequently, the specific ASC variant is represented by a three-component split Bragg peak, where the intensities of individual sub-peaks are determined by the volume ratios between the T-domains and the C-phase. The split peak patterns are schematically shown in Figure 4b.

The positions of the diffraction peaks are described by the reciprocal lattice vectors $\boldsymbol{B}_m = H\boldsymbol{a}_{1m}^* + K\boldsymbol{a}_{2m}^* + L\boldsymbol{a}_{3m}^*$, with $\boldsymbol{a}_{im}^*$ defined according to (A2.1). The separation of the Bragg peaks diffracted from T-domains $n$ and $m$ can be described by $\Delta\boldsymbol{B}_{mn} = \boldsymbol{B}_n - \boldsymbol{B}_m = \Delta H\boldsymbol{a}_{1m}^* + \Delta K\boldsymbol{a}_{2m}^* + \Delta L\boldsymbol{a}_{3m}^*$. As elaborated in Appendix A2:

$$[\Delta B]_{mn} = -[\Delta S]_{mn}^T \begin{pmatrix} H \\ K \\ L \end{pmatrix}, \qquad (32)$$

where, $[\Delta B]_{mn}$ is defined as a 3x1 column vector whose components are $\Delta H$, $\Delta K$, and $\Delta L$. According to (27) and (28):

$$[\Delta B]_{Cm,n} = [\Delta B]_{CA} + [\Delta B]_{Am,n} \qquad (33)$$

## 9. Summarizing table

The derivation of orientation relationships within the specific ASC $T_a(110)T_b$ variant was detailed above. It was demonstrated that an ASC connection exists when the volume ratios of the domains satisfy either $\omega_1 \approx \frac{x_c}{x_T}$ or $\omega_2 \approx \frac{\overline{x_a}}{x_T}$, with each case yielding two ASC variants. As summarized in Table 1, there are six T-domain pairs, and each pair generates four ASC variants. Consequently, there are 24 ASC variants, for which analogous derivation yields expressions similar to (18), (25), (26), (30), and (31).

Tables 3-14 summarize the result. Each table presents the relevant analytical expressions for two ASC variants. These tables include the HP "Miller" indices, the transformation matrices between the basis vectors, and the separations between the corresponding Bragg peaks, expressed as column vectors. To ensure brevity and clarity in the Tables, cells with identical values are merged. For example, the data in Table 3 can be used to calculate the separation of the Bragg peaks diffracted from the C phase and T-domain $n$ when ASC interface is parallel to the $(01\overline{M})$ plane according to

$$[\Delta B]_{Cn} = (x_a + x_c)(\overline{K} + M\overline{L})\begin{pmatrix} 0 \\ 1 \\ \overline{M} \end{pmatrix} + x_a(\overline{H} + \overline{K})\begin{pmatrix} 1 \\ \overline{1} \\ 0 \end{pmatrix}. \qquad (34)$$

Figure 4c shows a schematic diagram, supporting the definitions used in the Tables below. This figure shows the splitting of the Bragg peaks for the ASC variant corresponding to the $T_a(101)T_c$ domain pair and the HP parallel to the $(0M1)$ plane.

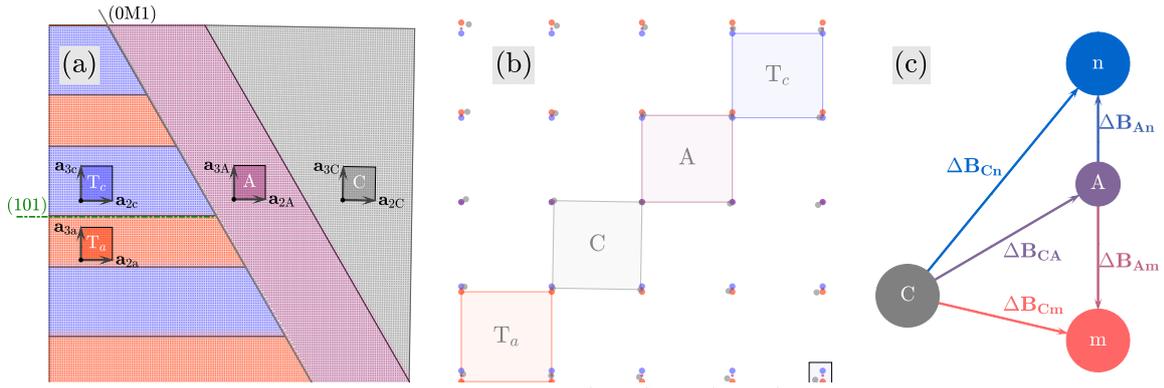

**Figure 4.** Schematic illustrations of the $(T_a(101)T_c)(0M1)C$ ASC variant, encompassing both real and reciprocal space drawings. **(a) Real space:** $0vw$ section of the crystal lattices, showing the tetragonal domains $T_a$ and $T_c$, the adaptive state (A), and the cubic phase (C). The plane $(101)$ – permissible domain wall between tetragonal domains – is inclined at 45° relative to the drawing plane, while the habit plane $(0M1)$ is perpendicular to it. **(b) Reciprocal space:** $0KL$ sections of the reciprocal lattices for all coexisting volumes, illustrating the concept of Bragg peak splitting. The reciprocal unit cells are separated for clarity. **(c) Bragg peak splitting:** A detailed depiction of the splitting pattern of the $0\bar{2}\bar{2}$ reflection (highlighted in b) from the specific ASC variant. The figure defines key vectors $\Delta \boldsymbol{B}_{CA}$, $\Delta \boldsymbol{B}_{Cm}$, $\Delta \boldsymbol{B}_{Cn}$, $\Delta \boldsymbol{B}_{Am}$, and $\Delta \boldsymbol{B}_{An}$ used in the text and in the Tables.

**Table 3.** ASC coexistence corresponding to the AS variant $T_a(1\bar{1}0)T_b$, $\omega_b = \omega_1 = \frac{x_c}{x_T}$. Here M is defined according to Equation (17). The first column lists the components of $[p]$. The second column shows $[\Delta S_{CA}]$, and the third column represents the separation between the corresponding Bragg peaks, defined by the column-vector $[\Delta B_{CA}]$. The next columns represent the same quantities for the orientation matrices $[\Delta S_{Am}]$ and $[\Delta S_{An}]$ and the corresponding separation between the Bragg peaks. The last row of the table shows the units of individual quantities above. For example, the data in this Table can be used to calculate the separation of the Bragg peaks between the C phase and b domain according to Equation (34).

| $[p]$ | $[\Delta S]_{CA}$ | $[\Delta B]_{CA}$ | $[\Delta S]_{Am}$ | $[\Delta S]_{An}$ | $[\Delta B]_{Am}$ | $[\Delta B]_{An}$ |
|---|---|---|---|---|---|---|
| $\begin{pmatrix}0\\1\\\bar{M}\end{pmatrix}$ | $\begin{pmatrix}0 & 0 & 0\\0 & 1 & \bar{M}\\0 & M & \bar{M}^2\end{pmatrix}$ | $(\bar{K}+M\bar{L})\begin{pmatrix}0\\1\\\bar{M}\end{pmatrix}$ | $\begin{pmatrix}1 & \bar{1} & 0\\1 & \bar{1} & 0\\0 & 0 & 0\end{pmatrix}$ | | $(\bar{H}+\bar{K})\begin{pmatrix}1\\\bar{1}\\0\end{pmatrix}$ | |
| $\begin{pmatrix}0\\1\\M\end{pmatrix}$ | $\begin{pmatrix}0 & 0 & 0\\0 & 1 & M\\0 & \bar{M} & \bar{M}^2\end{pmatrix}$ | $(\bar{K}+ML)\begin{pmatrix}0\\1\\M\end{pmatrix}$ | | | | |
| Units | $x_a + x_c$ | | $x_c$ | $x_a$ | $x_c$ | $x_a$ |

**Table 4.** The same as Table 3 but for the case of AS variant $T_a(1\bar{1}0)T_b$, $\omega_b = \omega_2 = \frac{\overline{x_a}}{x_T}$.

| $[p]$ | $[\Delta S]_{CA}$ | $[\Delta B]_{CA}$ | $[\Delta S]_{Am}$ | $[\Delta S]_{An}$ | $[\Delta B]_{Am}$ | $[\Delta B]_{An}$ |
|---|---|---|---|---|---|---|
| $\begin{pmatrix}1\\0\\\overline{M}\end{pmatrix}$ | $\begin{pmatrix}1&0&\overline{M}\\0&0&0\\M&0&\overline{M^2}\end{pmatrix}$ | $(\overline{H}+M\overline{L})\begin{pmatrix}1\\0\\\overline{M}\end{pmatrix}$ | $\begin{pmatrix}\bar{1}&1&0\\\bar{1}&1&0\\0&0&0\end{pmatrix}$ | | $(H+K)\begin{pmatrix}1\\\bar{1}\\0\end{pmatrix}$ | |
| $\begin{pmatrix}1\\0\\M\end{pmatrix}$ | $\begin{pmatrix}1&0&M\\0&0&0\\\overline{M}&0&\overline{M^2}\end{pmatrix}$ | $(\overline{H}+ML)\begin{pmatrix}1\\0\\M\end{pmatrix}$ | | | | |
| Units | $x_a + x_c$ | | $x_a$ | $x_c$ | $x_a$ | $x_c$ |

**Table 5.** The same as Table 3 but for the case of $T_a(110)T_b$, $\omega_b = \omega_1 = \frac{x_c}{x_T}$ connection.

| $[p]$ | $[\Delta S]_{CA}$ | $[\Delta B]_{CA}$ | $[\Delta S]_{Am}$ | $[\Delta S]_{An}$ | $[\Delta B]_{Am}$ | $[\Delta B]_{An}$ |
|---|---|---|---|---|---|---|
| $\begin{pmatrix}0\\1\\\overline{M}\end{pmatrix}$ | $\begin{pmatrix}0&0&0\\0&1&\overline{M}\\0&M&\overline{M^2}\end{pmatrix}$ | $(\overline{K}+M\overline{L})\begin{pmatrix}0\\1\\\overline{M}\end{pmatrix}$ | $\begin{pmatrix}1&1&0\\\bar{1}&\bar{1}&0\\0&0&0\end{pmatrix}$ | | $(\overline{H}+K)\begin{pmatrix}1\\1\\0\end{pmatrix}$ | |
| $\begin{pmatrix}0\\1\\M\end{pmatrix}$ | $\begin{pmatrix}0&0&0\\0&1&M\\0&\overline{M}&\overline{M^2}\end{pmatrix}$ | $(\overline{K}+ML)\begin{pmatrix}0\\1\\M\end{pmatrix}$ | | | | |
| Units | $x_a + x_c$ | | $x_c$ | $x_a$ | $x_c$ | $x_a$ |

**Table 6.** The same as Table 3 but for the case of $T_a(110)T_b$, $\omega_b = \omega_2 = \frac{\overline{x_a}}{x_T}$ connection.

| $[p]$ | $[\Delta S]_{CA}$ | $[\Delta B]_{CA}$ | $[\Delta S]_{Am}$ | $[\Delta S]_{An}$ | $[\Delta B]_{Am}$ | $[\Delta B]_{An}$ |
|---|---|---|---|---|---|---|
| $\begin{pmatrix}1\\0\\\overline{M}\end{pmatrix}$ | $\begin{pmatrix}1&0&\overline{M}\\0&0&0\\M&0&\overline{M^2}\end{pmatrix}$ | $(\overline{H}+M\overline{L})\begin{pmatrix}1\\0\\\overline{M}\end{pmatrix}$ | $\begin{pmatrix}\bar{1}&\bar{1}&0\\1&1&0\\0&0&0\end{pmatrix}$ | | $(H+\overline{K})\begin{pmatrix}1\\1\\0\end{pmatrix}$ | |
| $\begin{pmatrix}1\\0\\M\end{pmatrix}$ | $\begin{pmatrix}1&0&M\\0&0&0\\\overline{M}&0&\overline{M^2}\end{pmatrix}$ | $(\overline{H}+ML)\begin{pmatrix}1\\0\\M\end{pmatrix}$ | | | | |
| Units | $x_a + x_c$ | | $x_a$ | $x_c$ | $x_a$ | $x_c$ |

**Table 7.** The same as Table 3 but for the case of AS variant $T_a(10\bar{1})T_c$, $\omega_c = \omega_1 = \frac{x_c}{x_T}$.

| $[p]$ | $[\Delta S]_{CA}$ | $[\Delta B]_{CA}$ | $[\Delta S]_{Am}$ | $[\Delta S]_{An}$ | $[\Delta B]_{Am}$ | $[\Delta B]_{An}$ |
|---|---|---|---|---|---|---|
| $\begin{pmatrix} 0 \\ \bar{M} \\ 1 \end{pmatrix}$ | $\begin{pmatrix} 0 & 0 & 0 \\ 0 & M^2 & M \\ 0 & \bar{M} & 1 \end{pmatrix}$ | $(M\bar{K} + \bar{L})\begin{pmatrix} 0 \\ \bar{M} \\ 1 \end{pmatrix}$ | $\begin{pmatrix} 1 & 0 & \bar{1} \\ 0 & 0 & 0 \\ 1 & 0 & \bar{1} \end{pmatrix}$ | | $(\bar{H} + \bar{L})\begin{pmatrix} 1 \\ 0 \\ \bar{1} \end{pmatrix}$ | |
| $\begin{pmatrix} 0 \\ M \\ 1 \end{pmatrix}$ | $\begin{pmatrix} 0 & 0 & 0 \\ 0 & M^2 & \bar{M} \\ 0 & M & 1 \end{pmatrix}$ | $(MK + \bar{L})\begin{pmatrix} 0 \\ M \\ 1 \end{pmatrix}$ | | | | |
| Units | $x_a + x_c$ | | $x_c$ | $x_a$ | $x_c$ | $x_a$ |

**Table 8.** The same as Table 3 but for the case of AS variant $T_a(10\bar{1})T_c$, $\omega_c = \omega_2 = \frac{x_a}{x_T}$.

| $[p]$ | $[\Delta S]_{CA}$ | $[\Delta B]_{CA}$ | $[\Delta S]_{Am}$ | $[\Delta S]_{An}$ | $[\Delta B]_{Am}$ | $[\Delta B]_{An}$ |
|---|---|---|---|---|---|---|
| $\begin{pmatrix} 1 \\ \bar{M} \\ 0 \end{pmatrix}$ | $\begin{pmatrix} 1 & \bar{M} & 0 \\ M & M^2 & 0 \\ 0 & 0 & 0 \end{pmatrix}$ | $(\bar{H} + M\bar{K})\begin{pmatrix} 1 \\ \bar{M} \\ 0 \end{pmatrix}$ | $\begin{pmatrix} \bar{1} & 0 & 1 \\ 0 & 0 & 0 \\ \bar{1} & 0 & 1 \end{pmatrix}$ | | $(H + L)\begin{pmatrix} 1 \\ 0 \\ \bar{1} \end{pmatrix}$ | |
| $\begin{pmatrix} 1 \\ M \\ 0 \end{pmatrix}$ | $\begin{pmatrix} 1 & M & 0 \\ \bar{M} & M^2 & 0 \\ 0 & 0 & 0 \end{pmatrix}$ | $(\bar{H} + MK)\begin{pmatrix} 1 \\ M \\ 0 \end{pmatrix}$ | | | | |
| Units | $x_a + x_c$ | | $x_a$ | $x_c$ | $x_a$ | $x_c$ |

**Table 9.** The same as Table 3 but for the case of AS variant $T_a(101)T_c$, $\omega_c = \omega_1 = \frac{x_c}{x_T}$.

| $[p]$ | $[\Delta S]_{CA}$ | $[\Delta B]_{CA}$ | $[\Delta S]_{Am}$ | $[\Delta S]_{An}$ | $[\Delta B]_{Am}$ | $[\Delta B]_{An}$ |
|---|---|---|---|---|---|---|
| $\begin{pmatrix} 0 \\ \bar{M} \\ 1 \end{pmatrix}$ | $\begin{pmatrix} 0 & 0 & 0 \\ 0 & M^2 & M \\ 0 & \bar{M} & 1 \end{pmatrix}$ | $(M\bar{K} + \bar{L})\begin{pmatrix} 0 \\ \bar{M} \\ 1 \end{pmatrix}$ | $\begin{pmatrix} 1 & 0 & 1 \\ 0 & 0 & 0 \\ \bar{1} & 0 & \bar{1} \end{pmatrix}$ | | $(\bar{H} + L)\begin{pmatrix} 1 \\ 0 \\ 1 \end{pmatrix}$ | |
| $\begin{pmatrix} 0 \\ M \\ 1 \end{pmatrix}$ | $\begin{pmatrix} 0 & 0 & 0 \\ 0 & M^2 & \bar{M} \\ 0 & M & 1 \end{pmatrix}$ | $(MK + \bar{L})\begin{pmatrix} 0 \\ M \\ 1 \end{pmatrix}$ | | | | |
| Units | $x_a + x_c$ | | $x_c$ | $x_a$ | $x_c$ | $x_a$ |

Figure 5 provides a graphical illustration of the $T_a(101)T_c$ connection via the (0M1) HP.

**Table 10**. The same as Table 3 but for the case of AS variant $T_a(101)T_c$, $\omega_c = \omega_2$.

| $[p]$ | $[\Delta S]_{CA}$ | $[\Delta B]_{CA}$ | $[\Delta S]_{Am}$ | $[\Delta S]_{An}$ | $[\Delta B]_{Am}$ | $[\Delta B]_{An}$ |
|---|---|---|---|---|---|---|
| $\begin{pmatrix}1\\\bar{M}\\0\end{pmatrix}$ | $\begin{pmatrix}1 & \bar{M} & 0\\ M & \bar{M}^2 & 0\\ 0 & 0 & 0\end{pmatrix}$ | $(\bar{H}+M\bar{K})\begin{pmatrix}1\\\bar{M}\\0\end{pmatrix}$ | $\begin{pmatrix}\bar{1} & 0 & \bar{1}\\ 0 & 0 & 0\\ 1 & 0 & 1\end{pmatrix}$ | | $(H+\bar{L})\begin{pmatrix}1\\0\\1\end{pmatrix}$ | |
| $\begin{pmatrix}1\\M\\0\end{pmatrix}$ | $\begin{pmatrix}1 & M & 0\\ \bar{M} & \bar{M}^2 & 0\\ 0 & 0 & 0\end{pmatrix}$ | $(\bar{H}+MK)\begin{pmatrix}1\\M\\0\end{pmatrix}$ | | | | |
| Units | $x_a + x_c$ | | $x_a$ | $x_c$ | $x_a$ | $x_c$ |

**Table 11**. The same as Table 3 but for the case of AS variant $T_b(01\bar{1})T_c$, $\omega_c = \omega_1$.

| $[HP]$ | $[\Delta S]_{CA}$ | $[\Delta B]_{CA}$ | $[\Delta S]_{Am}$ | $[\Delta S]_{An}$ | $[\Delta B]_{Am}$ | $[\Delta B]_{An}$ |
|---|---|---|---|---|---|---|
| $\begin{pmatrix}\bar{M}\\0\\1\end{pmatrix}$ | $\begin{pmatrix}\bar{M}^2 & 0 & M\\ 0 & 0 & 0\\ \bar{M} & 0 & 1\end{pmatrix}$ | $(M\bar{H}+\bar{L})\begin{pmatrix}\bar{M}\\0\\1\end{pmatrix}$ | $\begin{pmatrix}0 & 0 & 0\\ 0 & 1 & \bar{1}\\ 0 & 1 & \bar{1}\end{pmatrix}$ | | $(\bar{K}+\bar{L})\begin{pmatrix}0\\1\\\bar{1}\end{pmatrix}$ | |
| $\begin{pmatrix}M\\0\\1\end{pmatrix}$ | $\begin{pmatrix}\bar{M}^2 & 0 & \bar{M}\\ 0 & 0 & 0\\ M & 0 & 1\end{pmatrix}$ | $(MH+\bar{L})\begin{pmatrix}M\\0\\1\end{pmatrix}$ | | | | |
| Units | $x_a + x_c$ | | $x_c$ | $x_a$ | $x_c$ | $x_a$ |

**Table 12**. The same as Table 3 but for the case AS variant $T_b(01\bar{1})T_c$, $\omega_c = \omega_2$ connection.

| $[p]$ | $[\Delta S]_{CA}$ | $[\Delta B]_{CA}$ | $[\Delta S]_{Am}$ | $[\Delta S]_{An}$ | $[\Delta B]_{Am}$ | $[\Delta B]_{An}$ |
|---|---|---|---|---|---|---|
| $\begin{pmatrix}\bar{M}\\1\\0\end{pmatrix}$ | $\begin{pmatrix}\bar{M}^2 & M & 0\\ \bar{M} & 1 & 0\\ 0 & 0 & 0\end{pmatrix}$ | $(M\bar{H}+\bar{K})\begin{pmatrix}\bar{M}\\1\\0\end{pmatrix}$ | $\begin{pmatrix}0 & 0 & 0\\ 0 & \bar{1} & 1\\ 0 & \bar{1} & 1\end{pmatrix}$ | | $(K+L)\begin{pmatrix}0\\1\\\bar{1}\end{pmatrix}$ | |

| [p] | [ΔS]$_{CA}$ | [ΔB]$_{CA}$ | | | |
|---|---|---|---|---|---|
| $\begin{pmatrix}M\\1\\0\end{pmatrix}$ | $\begin{pmatrix}\overline{M}^2 & \overline{M} & 0\\ M & 1 & 0\\ 0 & 0 & 0\end{pmatrix}$ | $(MH+\overline{K})\begin{pmatrix}M\\1\\0\end{pmatrix}$ | | | |
| Units | $x_a + x_c$ | | $x_a$ | $x_c$ | $x_a$ | $x_c$ |

**Table 13.** The same as Table 3 but for the case of AS variant $T_b(011)T_c$, $\omega_c = \omega_1$.

| [p] | [ΔS]$_{CA}$ | [ΔB]$_{CA}$ | [ΔS]$_{Am}$ | [ΔS]$_{An}$ | [ΔB]$_{Am}$ | [ΔB]$_{An}$ |
|---|---|---|---|---|---|---|
| $\begin{pmatrix}\overline{M}\\0\\1\end{pmatrix}$ | $\begin{pmatrix}\overline{M}^2 & 0 & M\\ 0 & 0 & 0\\ \overline{M} & 0 & 1\end{pmatrix}$ | $(M\overline{H}+\overline{L})\begin{pmatrix}\overline{M}\\0\\1\end{pmatrix}$ | $\begin{pmatrix}0 & 0 & 0\\ 0 & 1 & 1\\ 0 & \overline{1} & \overline{1}\end{pmatrix}$ | | $(\overline{K}+L)\begin{pmatrix}0\\1\\1\end{pmatrix}$ | |
| $\begin{pmatrix}M\\0\\1\end{pmatrix}$ | $\begin{pmatrix}\overline{M}^2 & 0 & \overline{M}\\ 0 & 0 & 0\\ M & 0 & 1\end{pmatrix}$ | $(MH+\overline{L})\begin{pmatrix}M\\0\\1\end{pmatrix}$ | | | | |
| Units | $x_a + x_c$ | | $x_c$ | $x_a$ | $x_c$ | $x_a$ |

**Table 14.** The same as Table 3 but for the case of AS variant $T_b(011)T_c$, $\omega_c = \omega_2$.

| [p] | [ΔS]$_{CA}$ | [ΔB]$_{CA}$ | [ΔS]$_{Am}$ | [ΔS]$_{An}$ | [ΔB]$_{Am}$ | [ΔB]$_{An}$ |
|---|---|---|---|---|---|---|
| $\begin{pmatrix}\overline{M}\\1\\0\end{pmatrix}$ | $\begin{pmatrix}\overline{M}^2 & M & 0\\ \overline{M} & 1 & 0\\ 0 & 0 & 0\end{pmatrix}$ | $(M\overline{H}+\overline{K})\begin{pmatrix}\overline{M}\\1\\0\end{pmatrix}$ | $\begin{pmatrix}0 & 0 & 0\\ 0 & \overline{1} & \overline{1}\\ 0 & 1 & 1\end{pmatrix}$ | | $(K+\overline{L})\begin{pmatrix}0\\1\\1\end{pmatrix}$ | |
| $\begin{pmatrix}M\\1\\0\end{pmatrix}$ | $\begin{pmatrix}\overline{M}^2 & \overline{M} & 0\\ M & 1 & 0\\ 0 & 0 & 0\end{pmatrix}$ | $(MH+\overline{K})\begin{pmatrix}M\\1\\0\end{pmatrix}$ | | | | |
| Units | $x_a + x_c$ | | $x_a$ | $x_c$ | $x_a$ | $x_c$ |

## 10. Experimental verification of the derived orientation relationship.

### 10.1 High-resolution reciprocal space mapping

The methodology for high-resolution reciprocal space mapping using a monochromatic X-ray beam, a four-circle goniometer and a pixel area detector has been extensively detailed in [28,29,34]. This experimental approach enables the reconstruction of X-ray diffraction intensity distributions around selected Bragg peaks to resolve the separations between nearly overlapping peaks components, each arising from a distinct domain variant. These experiments have been made possible by the recent advancement in synchrotron-based and laboratory X-ray sources, the development of pixel area detectors, the improved beam conditioning systems and the enhanced big-data exchange protocols, as outlined in [30,44,45].

## 10.2 Recognition of ASC variants at the C-T phase transition in PMN-35PT.

This paragraph demonstrates the application of the above-derived formalisms to the case of CT phase transition in relaxor-based piezo-/ferroelectric $0.65PbMg_{2/3}Nb_{1/3}O_3 - 0.35PbTiO_3$ (PMN − 35PT) crystals. (1-x)PMN- x PT is a perovskite-based solid solution that has attracted interest due to its exceptional electromechanical properties [46]. The phase diagram of PMN-PT features a transition from a high-temperature cubic paraelectric phase to a low-temperature tetragonal (for x>0.35) or rhombohedral (for x<0.28) phase [6,47,48]. The cubic-to-tetragonal transition in PMN-35PT at approximately 435 K is used in this study as a case example.

X-ray diffraction experiment was conducted at the ID28 beamline of the European Synchrotron Radiation facility (ESRF), as described in [49]. Using the CrysAlisPro software [50], we got the orientation matrix averaged over all the domains in the X-ray beam. Then we acquired temperature-dependent high-resolution reciprocal space maps of the diffraction intensity distribution around several Bragg peaks. The collected data were then represented as intensity tables $I(B_x, B_y, B_z)$ where $B_x, B_y,$ and $B_z$ are the components of the scattering vector relative to the Cartesian axes, **X**, **Y**, and **Z**, with the **X**-axis aligned to the scattering vector ***B***. Figures 6 (*a, b, c*) and 7 (*a, b, c*) show the $B_xB_y, B_xB_z, B_yB_z$ projections of the three-dimensional X-ray intensity distribution $I(B_x, B_y, B_z)$.

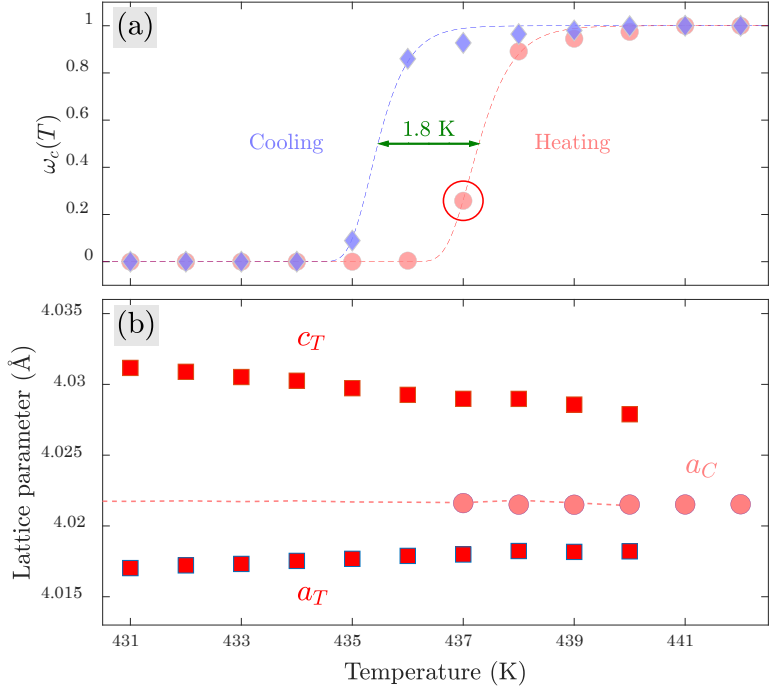

**Figure 5. Experimental results adapted from** [49].

(a) Temperature dependence of the C phase fraction during heating and cooling across the CT transition in PMN-35PT. The highlighted data point (Large circle) corresponds to the one featured in Figure 6 - 9.

(b) Temperature dependence of the T and C lattice parameters, shown both within and outside the phase coexistence region.

The temperature dependences of 3D-reciprocal space maps have been published in [49]. For the convenience of the reader, we provide these temperature dependences around the phase transition in the Supplemental Materials. These materials are available in the form of a single power point presentation file with embedded animated temperature dependences of reciprocal space maps. Each of such reciprocal space maps highlights the multi-component nature of the Bragg peak, with all the peak components merging in the C-phase. The continuity of these merged peaks position enables straightforward assignment of the sub-peaks diffracted from the C phase and facilitates the estimation of the C-phase fraction within the total crystal volume in the X-ray beam.

Figure 5a uses the data from [49] and illustrates the CT phase coexistence over a 4K temperature range. It shows the temperature dependence of the C-phase volume fraction. Figure 5b shows that of $a_C, a_T$ and $c_T$, highlighting that $a_C = \frac{2}{3}a_T + \frac{1}{3}c_T$ ($x_c = -2x_a$). According to Table 2, this yields M = 1 and sets HP along one of the {110} lattice planes. The connection is possible for two domain ratios $\omega_1 = \frac{x_c}{x_T} = \frac{2}{3}$ and $\omega_2 = \frac{\overline{x_a}}{x_T} = \frac{1}{3}$. Figures 6-9 display the projections of the 3D diffraction intensity distribution with the added separation between the Bragg peaks, each according to one Tables 3-14. These projections demonstrate a strong agreement between the predicted and experimentally observed Bragg peak separations, thereby validating the predicted ASC orientation relationships.

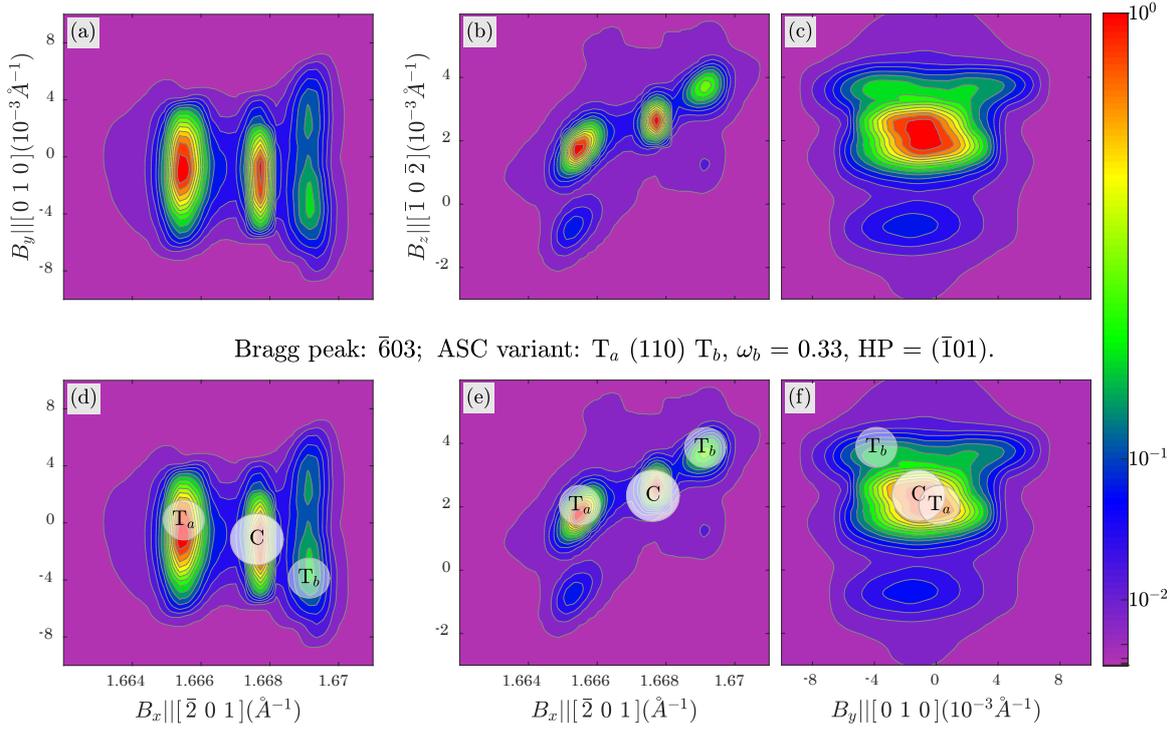

**Figure 6.** Three-dimensional diffraction intensity distribution around the split family of peaks with the indices $\bar{6}03$ of PMN-35PT measured at the C − T phase transition at 437 K. Panels (a-c) show the projections of the diffraction intensity distribution $I_z(B_x B_y), I_y(B_x B_z), I_x(B_x B_y)$, respectively. The individual components of the split peaks arise from the "C" and "T" regions of the crystal, with the T-regions further giving the contribution to the peak splitting. Panels (d-f) display the same intensity distributions, with the positions of the sub-peaks modeled according to the ASC variant. This variant is formed by the $T_a(110)T_b$ $\omega_b = \omega_2$ domain assemblage, connected to the C-phase along the Habit Plane $(\bar{1}01)$. This ASC - variant is shown in Table 6.

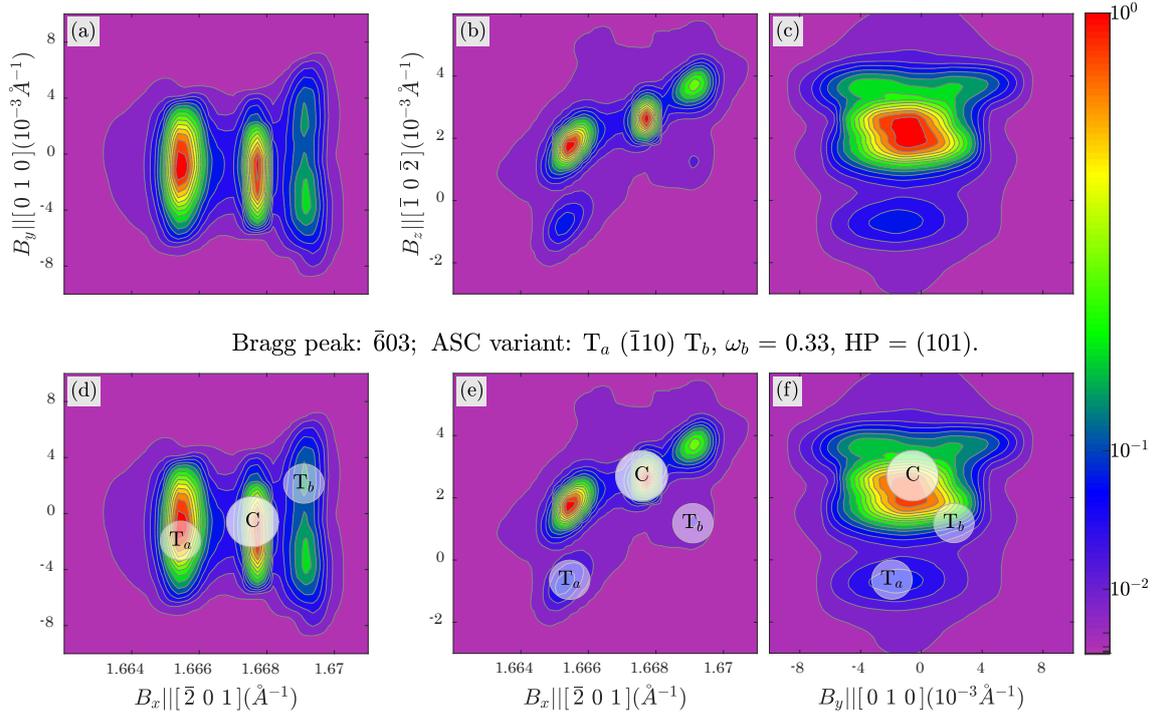

**Figure 7.** The same as Figure 6 but with the sub-peaks modelled by the AS formed by the $T_a(1\bar{1}0)T_b$ $\omega_b = \omega_2$ connected to the coexisting C - phase along the Habit Plane (101). This ASC – variant is shown in Table 4.

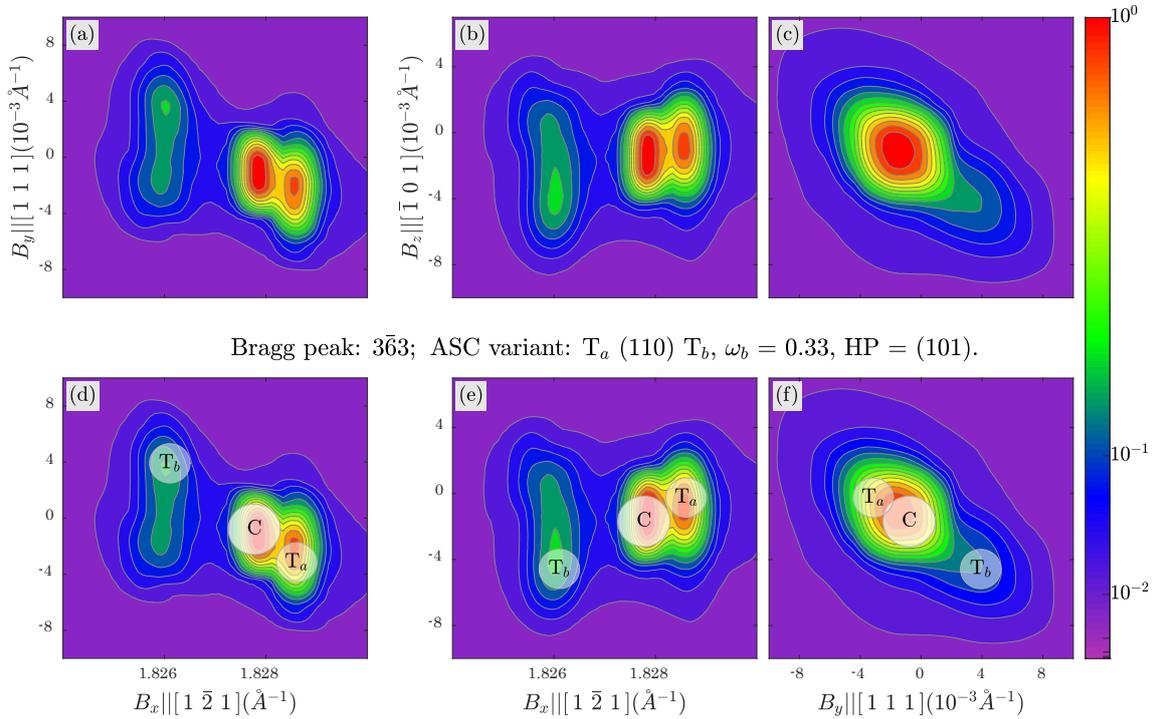

**Figure 8.** Three-dimensional diffraction intensity distribution around the split family of peaks with the indices $3\bar{6}3$ measured at 437 K. The figure is organized as Figures 6 and 7 and involves the modelling of the Bragg peak splitting predicted for the AS-variant $T_a(110)T_b$ $\omega_b = \omega_2$ and the Habit Plane (101). This variant is shown in Table 6.

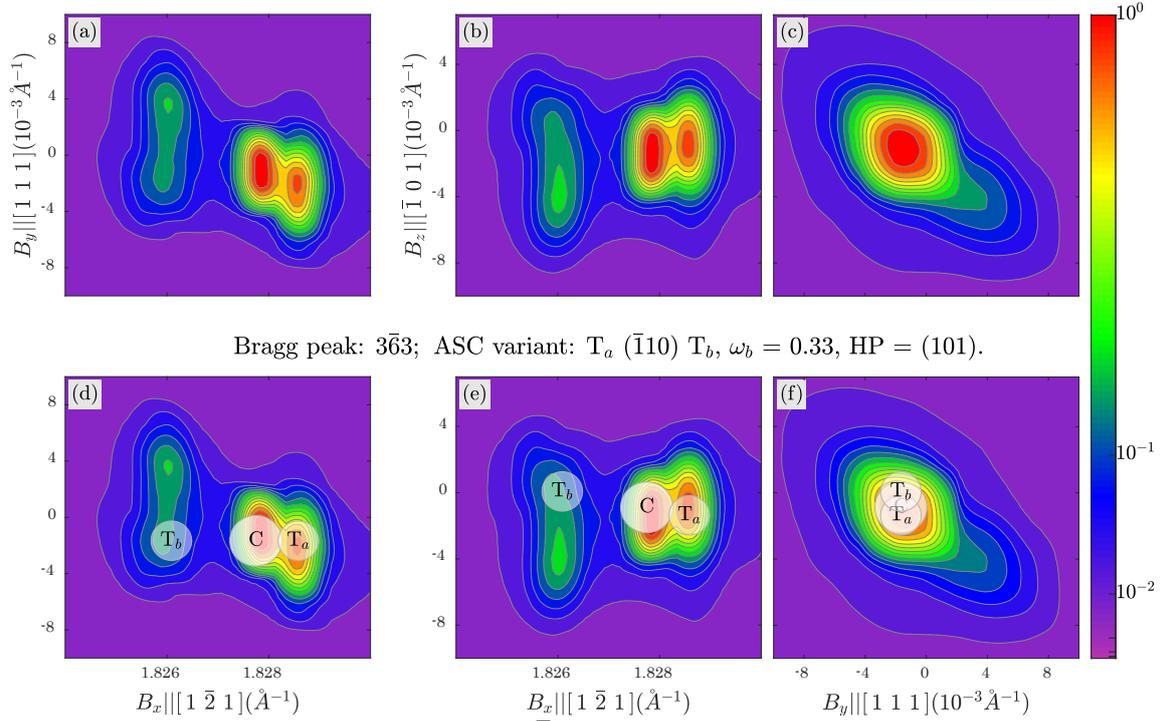

**Figure 9.** The same as Figure 8 but $T_a(1\bar{1}0)T_b$ $\omega_b = \omega_2$ connected to the coexisting C - phase along the Habit Plane (101). This ASC variant is shown in Table 4.

## 11. Discussion

We have investigated the mismatch-free connections between the coexisting C and T phases along a planar interface, commonly referred to as a Habit Plane (HP). Such an interface cannot form between pure C and T phases; instead, it is enabled through the "macroscopic averaging" of the constituent ferroelastic domains. In this approach, the multi-domain state of the T-phase is approximated as a uniform "single-domain" state, with its virtual lattice basis vectors averaged from the "basis vectors" of the constituent T-domains. This averaging introduces an additional parameter, $\omega$, representing the volume ratio between the T-domains in the assemblage. Although $\omega$ is a microstructural rather than purely crystallographic parameter, it affects the virtual lattice parameters of the average crystal and enables their **adaptation** to the conditions necessary for matching with the coexisting C - phase.

The concept has been discussed in several earlier publications on ferroelastic materials [38,39]. The specific aim of our work is to derive analytical expressions that describe the splitting of Bragg peaks diffracted from all participating domains and phases. These results are expected to be valuable for analyzing high-resolution X-ray and electron diffraction patterns. To this end, we have developed analytical expressions describing the crystallographic relationships between the C-, AS-, and T-phases. These relationships are not only central to our diffraction analysis but are also useful for broader characterization of the resulting ASC microstructure.

Previous work by Wang [42,43], examined X-ray diffraction from AS-like systems, focusing on single-phase structures containing domains of a single symmetry (e.g., tetragonal). Wang also discussed the effects of domain miniaturization on the diffraction signal. However, those studies did not address systems involving multiple coexisting phases. In contrast, our work generalizes Wang's formalism by predicting the full Bragg peak splitting pattern arising from all contributing phases and domain states.

The formation of the AS phase through macroscopic averaging inherently assumes domain miniaturization. That is, the size of individual domains in the direction perpendicular to the domain walls (see Figures 1 and 2) is much smaller than the overall extent of the domain assemblage. Notably, these domain sizes can vary over three orders of magnitude, ranging from approximately 0.5 to 500 nm. Measuring the actual domain dimensions is challenging, especially since domain size can vary significantly across the crystal and depends sensitively on temperature near the phase transition.

Our approach enables the modelling of Bragg peaks under the assumption that the orientational relationship between the phases follow those expected from the AS-like mechanism of coexistence. We further assume that the individual T-domains are large enough to produce independent diffraction peaks - an assumption fully validated by the X-ray diffraction results, presented in Figures 6-9. Assuming that domain size is the dominant contribution to the observed Bragg peaks widths (0.002–0.005 $\text{Å}^{-1}$), we estimate the domain size to be several hundred nanometers. This result does not preclude the presence of domain miniaturization effects, which may occur closer to the phase transition. Nonetheless, we show that the orientation relationship between the C-phase and the AS state remains preserved, even when T-domains become larger.

Beyond guiding and interpreting X-ray diffraction experiments, our results may be valuable for a broader range of problems where the orientation relationships between coexisting phases play a critical role. Future research may focus on applying these methods to other ferroelectric systems, potentially uncovering new insights into phase transitions and domain behaviors across a wider range of materials. Additionally, a similar method can be applied to the description of coexistence of other phases, e.g. the rhombohedral and tetragonal phases which are commonly present in the morphotropic phase boundary of many piezo-/ferroelectric solid solution systems.

## 12. Conclusions

In this work, we have developed formalisms for the use of single-crystal X-ray diffraction techniques to investigate domain microstructures during phase transitions in ferroelectric materials. By extending the Weschler—Lieberman—Read (WLR) crystallographic theory and the concept of adaptive state (AS), we have derived the analytical expressions for the orientation relationships between the real and reciprocal lattices of the cubic phase and tetragonal domains at the phase transition for 24 possible variants of their mismatch-free connections. We have successfully implemented the developed formalisms for the analysis of the Bragg peaks from the high-resolution reciprocal space maps of the perovskite-based PMN-35PT single crystal at phase coexistence temperatures, demonstrating the validity of the derived twinning relationship in the coexisting cubic and tetragonal phases of ferroelectrics.

**Appendix A. Important definitions and crystallographic relationship**

This paper uses the following crystallographic abbreviations, notations and relationships. Throughout all the paper C stands for "**cubic**", T stands "**tetragonal**", AS stands for "**adaptive state**"

**A1: Real space**

**C lattice parameter** $a_C$ defines the lattice period along four-fold symmetry axes in the cubic $Pm\bar{3}m$ phase. For perovskite oxides, $a_C$ is close to 4 Å.

**T lattice parameters** $a_T, c_T$ define the lengths of the "pseudo-cubic unit cell" edges in the tetragonal (e.g. $P4mm$) phase. We adopt the common definition of the "pseudo-cubic unit cell," where the edges are ~4 Å long, and all angles are approximately 90°.

**The "pseudo-cubic" basis vectors** $\boldsymbol{a}_{1m}, \boldsymbol{a}_{2m}, \boldsymbol{a}_{3m}$ are defined by the edges of the "pseudo-cubic unit cell". These vectors can be obtained by a small distortion of the corresponding basis vectors in the cubic phase. Here $m = $ a, b, c for the three T domains, $m = $ C for the C-phase, $m = $ A for the case when these vectors are defined by averaging within a domain assemblage.

It is important to note that the "pseudo-cubic unit cell" may differ from the true crystallographic unit cell. A **transition** to a lower-symmetry phase can lead to doubling of the unit cell dimensions, changes in the symmetry of the crystal lattice or / and alternations in the conventional unit cell setting. This change is governed by specific transformation laws. Nevertheless, the "pseudocubic basis vectors" provide convenient reference coordinate system and are often used for the description of structures and indexing diffraction pattern.

**T distortion parameters** $x_a, x_c$ define the following dimensionless quantities:

$$x_a = \frac{a_T}{a_C} - 1,$$
$$x_c = \frac{c_T}{a_C} - 1. \tag{A1.1}$$

It is common that $x_a < 0$ and $x_c > 0$. In addition, $x_T, \tau$ are defined as:

$$x_T = \frac{c_T}{a_T} - 1, \tag{A1.2}$$

$$\tau = \frac{c_T^2 - a_T^2}{c_T^2 + a_T^2}. \tag{A1.3}$$

It is easy to verify that $\tau = x_T + O(x_T^2)$.

**The matrix of dot product** $[G]_m$ is $G_{ij,m} = (\boldsymbol{a}_{im} \cdot \boldsymbol{a}_{jm})$, $m = $ a, b, c, C, A. For the C phase,

$$[G]_C = \begin{pmatrix} a_C^2 & 0 & 0 \\ 0 & a_C^2 & 0 \\ 0 & 0 & a_C^2 \end{pmatrix} = a_C^2[I] \ . \tag{A1.4}$$

For the T phase and domain variants $T_a$, $T_b$, $T_c$:

$$[G]_a = \begin{pmatrix} c_T^2 & 0 & 0 \\ 0 & a_T^2 & 0 \\ 0 & 0 & a_T^2 \end{pmatrix}, \ [G]_b = \begin{pmatrix} a_T^2 & 0 & 0 \\ 0 & c_T^2 & 0 \\ 0 & 0 & a_T^2 \end{pmatrix}, \ [G]_c = \begin{pmatrix} a_T^2 & 0 & 0 \\ 0 & a_T^2 & 0 \\ 0 & 0 & c_T^2 \end{pmatrix}. \tag{A1.5}$$

**Transformation between the basis vectors** $\boldsymbol{a}_{im} \to \boldsymbol{a}_{in}$ is described by the 3x3 matrix $[S]_{mn}$ (here $n, m = $ a, b, c, C, A):

$$(\boldsymbol{a}_{1n} \quad \boldsymbol{a}_{2n} \quad \boldsymbol{a}_{3n}) = (\boldsymbol{a}_{1m} \quad \boldsymbol{a}_{2m} \quad \boldsymbol{a}_{3m})[S]_{mn} \ . \tag{A1.6}$$

Equation (A1.6) means that the columns of $[S]_{mn}$ are the coordinates of the vectors $\boldsymbol{a}_{1n}, \boldsymbol{a}_{2n}, \boldsymbol{a}_{3n}$ relative to $\boldsymbol{a}_{1m}, \boldsymbol{a}_{2m}, \boldsymbol{a}_{3m}$. Additionally, we define "delta" transformation matrix $[\Delta S]_{mn}$:

$$[\Delta S]_{mn} = [S]_{mn} - [I] \ . \tag{A1.7}$$

**The following relationship** between $[G]_m$ and $[G]_n$ exists:

$$[G]_n = [S]_{mn}^T [G]_m [S]_{mn} \ . \tag{A1.8}$$

**Two-step transformation** describes the transformation $\boldsymbol{a}_{im} \to \boldsymbol{a}_{in}$ via the intermediate step $\boldsymbol{a}_{ip}$. If $\boldsymbol{a}_{im} \to \boldsymbol{a}_{ip}$ is defined by $[S]_{mp}$ and $\boldsymbol{a}_{ip} \to \boldsymbol{a}_{in}$ is defined by $[S]_{pn}$, then $\boldsymbol{a}_{im} \to \boldsymbol{a}_{in}$ is given by

$$[S]_{mn} = [S]_{mp}[S]_{pn} \ . \tag{A1.9}$$

Neglecting the second power term $[\Delta S]_{mp}[\Delta S]_{pn}$, we can re-write (A1.9) as:

$$[\Delta S]_{mn} \approx [\Delta S]_{mp} + [\Delta S]_{pn} \ . \tag{A1.10}$$

**A2: Reciprocal space**

**Reciprocal "pseudo-cubic" basis vectors**: $\boldsymbol{a}_{1m}^*, \boldsymbol{a}_{2m}^*, \boldsymbol{a}_{3m}^*$ are defined using the following relationship to their real space counterparts $\boldsymbol{a}_{1m}, \boldsymbol{a}_{2m}, \boldsymbol{a}_{3m}$:

$$\boldsymbol{a}_{im}^* \boldsymbol{a}_{jm} = \delta_{ij} \ . \tag{A2.1}$$

**The matrix of reciprocal dot product** $[G^*]_m$ is defined as $G_{ij,m}^* = (\boldsymbol{a}_{im}^* \boldsymbol{a}_{jn}^*)$. Importantly, the following relationship between $[G^*]_m$ and $[G]_m$ exists:

$$[G^*]_m = [G]_m^{-1} \ . \tag{A2.2}$$

**Transformation between reciprocal basis vectors** $\boldsymbol{a}_{im}^* \to \boldsymbol{a}_{in}^*$ $[S^*]_{mn}$ defines the reciprocal basis vectors $\boldsymbol{a}_{in}^*$ relative to $\boldsymbol{a}_{im}^*$ in the way, identical to (A1.6), so that:

$$(\boldsymbol{a}^*_{1n} \quad \boldsymbol{a}^*_{2n} \quad \boldsymbol{a}^*_{3n}) = (\boldsymbol{a}^*_{1m} \quad \boldsymbol{a}^*_{2m} \quad \boldsymbol{a}^*_{3m})[S^*]_{mn}. \tag{A2.3}$$

It is easy to show that if the transformation $\boldsymbol{a}_{im} \to \boldsymbol{a}_{in}$ is described by the matrix $[S_{mn}]$ then:

$$[S^*]^T_{mn} = [S]^{-1}_{mn}. \tag{A2.4}$$

Additionally, this paper uses delta transformation matrix $[\Delta S^*]_{mn}$ defined according to

$$[\Delta S^*]_{mn} = [S^*]_{mn} - [I]. \tag{A2.5}$$

Using (A2.4) we can write

$$([I] + [\Delta S^*]^T_{mn})([I] + [\Delta S]_{mn}) = [I]. \tag{A2.6}$$

Neglecting the term $[\Delta S^*]^T_{mn}[\Delta S]_{mn}$ we can derive the following relationship:

$$[\Delta S^*]^T_{mn} \approx -[\Delta S]_{mn}. \tag{A2.7}$$

**Indices of reflection** $H, K, L$ define the coordinates of a reciprocal lattice vector $\boldsymbol{B}$ with respect to the reciprocal pseudocubic basis vectors. Specifically, for the domain variant $m$: $\boldsymbol{B}_m = H\boldsymbol{a}^*_{1m} + K\boldsymbol{a}^*_{2m} + L\boldsymbol{a}^*_{3m}$.

**The reciprocal space separation** of a Bragg peak diffracted from domains $n$ and $m$ is $\Delta \boldsymbol{B}_{mn} = \boldsymbol{B}_n - \boldsymbol{B}_m$ and the components $\Delta \boldsymbol{B}_{mn} = \Delta H \boldsymbol{a}^*_{1m} + \Delta K \boldsymbol{a}^*_{2m} + \Delta L \boldsymbol{a}^*_{3m}$. It is straightforward to see that following the definition (A2.3):

$$\begin{pmatrix} \Delta H \\ \Delta K \\ \Delta L \end{pmatrix} = [\Delta S^*]_{mn} \begin{pmatrix} H \\ K \\ L \end{pmatrix}. \tag{A2.8}$$

**Appendix B. Mismatch-free connection between various domains and phases**

**B1: The indices of the planar interface**

The mathematical formalism for analyzing the mismatch-free connection between various domains and phases has been explained in detail in our previous works [34,35]. For the reader's convenience, we summarize the key steps here. Importantly the algorithm is valid for the case of the same phase ferroelastic domains and for e.g. C phase and AS. Assuming that the connected volumes have the matrices of dot products $[G]_C$ and $[G]_A$, the mismatch free interface between them should satisfy the condition:

$$G_{ij,A} x_i x_j = G_{ij,C} x_i x_j, \tag{B1.1}$$

where $i, j = 1, 2, 3$, and $x_1, x_2, x_3$ are the coordinates of the points in the interface, relative to the pseudo-cubic basis. For this interface to be a plane, it is necessary that:

$$h_1 x_1 + h_2 x_2 + h_3 x_3 = 0, \tag{B1.2}$$

where $h_1, h_2, h_3$ are the "Miller" indices of the interface plane. Also, $h_i$ are the coordinates of the interface normal relative to the reciprocal pseudocubic basis. (B1.1) can be reduced to (B1.2) via the following steps:

**Step 1**: Transforming $[\Delta G]_{AC} = [G]_A - [G]_C$ to the diagonal form. This can be done by finding three eigenvalues $\lambda_1$, $\lambda_2$, $\lambda_3$. Introducing the matrix $[V]$ whose columns are the corresponding eigenvectors $\mathbf{v}_1, \mathbf{v}_2, \mathbf{v}_3$ of $[\Delta G]_{AC}$ and the following transformation (the Einstein summation over the repeated index is implemented here):

$$x_i = V_{ik} v_k , \qquad (B1.3)$$

we can re-write (B1.1) in the following form:

$$\lambda_1 v_1^2 + \lambda_2 v_2^2 + \lambda_3 v_3^2 = 0 . \qquad (B1.4)$$

**Step 2**: Checking if at least one of the eigenvalues is zero while two others have opposite signs. For certainty we assume $\lambda_2 = 0$, $\lambda_1 < 0$ and $\lambda_3 > 0$ and re-write (B1.4) as:

$$\Lambda^2 v_1^2 - v_3^2 = 0 , \qquad (B1.5)$$

where

$$\Lambda = \left(\frac{|\lambda_1|}{\lambda_3}\right)^{\frac{1}{2}} . \qquad (B1.6)$$

This leads to two different mismatch-free planar interfaces given by:

$$\begin{matrix} \Lambda v_1 - v_3 = 0 \\ \Lambda v_1 + v_3 = 0 \end{matrix} . \qquad (B1.7)$$

The orthogonality of $[V]$ ($[V]^{-1} = [V]^T$) means that (B1.3) can be re-written as

$$v_i = V_{ki} x_k . \qquad (B1.8)$$

And accordingly, the equations (B1.7) take the form:

$$\begin{matrix} (\Lambda V_{i1} - V_{i3}) x_i = 0 \\ (\Lambda V_{i1} + V_{i3}) x_i = 0 \end{matrix} , \qquad (B1.9)$$

so that the "Miller indices" of the mismatch-free planar interface are given by:

$$\begin{matrix} p_{i,1} = (\Lambda V_{i1} - V_{i3}) \\ p_{i,2} = (\Lambda V_{i1} + V_{i3}) \end{matrix} . \qquad (B1.10)$$

The procedure described above can be readily implemented in a computer program. However, in many cases $[p]_{1,2}$ can also be derived analytically. This has been demonstrated for the tetragonal [34], rhombohedral [34], monoclinic $M_A$ and $M_B$ [35] and monoclinic $M_C$ domains [36]. Paragraph 5 illustrates the derivation of the HP indices for the connection between the $T_a(110)T_b$ AS variant and the C phase. It reveals that such a connection is feasible when the volume fraction of the $T_b$ domain in the assemblage equals to either $\omega_1 = \frac{x_c}{x_T}$ or $\omega_2 = \frac{\bar{x}_a}{x_T}$. Both cases yield

$$\Lambda \approx M, \; M = \left(\frac{\bar{x}_a}{x_a + x_c}\right)^{\frac{1}{2}} . \qquad (B1.11)$$

For $\omega_1$ it is shown that two HP with "Miller" indices $(0\bar{1}M)$ and $(01M)$ are possible.

**B2: The relationship between the basis vectors of the connected volumes**

The formalism for finding the Miller indices of the mismatch-free planar interfaces can be extended for the derivation of the transformation relationship defined in (A1.6) and (A1.7). Such transformation can then be expressed by the matrix $[\Delta S]_{CA}$. Finding this matrix involves the following steps:

**Step 1:** Calculating the elements of the matrix:

$$[Z] = \begin{pmatrix} 1 & 0 & 0 \\ 0 & 1 & 0 \\ Z_{31} & 0 & \Lambda \end{pmatrix}, \tag{B2.1}$$

where $Z_{31} = +\Lambda$ and $Z_{31} = -\Lambda$ correspond to the cases of the interfaces normal to $\boldsymbol{p}_1$ and $\boldsymbol{p}_2$, respectively.

**Step 2:** Calculating the elements of $[G']_C$ and $[G']_A$:

$$[G']_m = [Z]^T [V]^T [G]_m [V][Z]. \tag{B2.2}$$

These are the matrices of dot products between the transformed basis vectors $\mathbf{w}_1, \mathbf{w}_2, \mathbf{w}_3$ such that $\mathbf{w}_1 = \mathbf{v}_1 + Z_{31}\mathbf{v}_3$, $\mathbf{w}_2 = \mathbf{v}_2$, $\mathbf{w}_3 = \mathbf{v}_3$. Importantly, this coordinate system allows us to describe the habit planes in (B1.10) as $(001)_w$.

**Step 3:** Calculating the coefficient $y_3$ as the ratio of the determinants according to (*20*)

$$y_3 = \left(\frac{|G|_A}{|G|_C}\right)^{\frac{1}{2}} \approx 1 + 2x_a + x_c, \tag{B2.3}$$

and $y_1$ and $y_2$ according to:

$$\begin{pmatrix} G'_{11,C} & G'_{12,C} \\ G'_{21,C} & G'_{22,C} \end{pmatrix} \begin{pmatrix} y_1 \\ y_2 \end{pmatrix} = \begin{pmatrix} G'_{13,A} - G'_{13,C} y_3 \\ G'_{23,A} - G'_{23,C} y_3 \end{pmatrix}. \tag{B2.4}$$

**Step 4:** Defining $[S']_{CA}$ for the orientation relationship $\mathbf{w}_{iC} \to \mathbf{w}_{iA}$:

$$[S']_{CA} = \begin{pmatrix} 1 & 0 & y_1 \\ 0 & 1 & y_2 \\ 0 & 0 & y_3 \end{pmatrix} \tag{B2.5}$$

The shape of this transformation matrix expresses the fact that the HP is parallel to $(001)_w$.

**Step 5:** Obtaining the transformation matrix $[S]_{CA}$ using

$$[S]_{CA} = [V][Z][S']_{CA}[Z]^{-1}[V]^T, \tag{B2.6}$$

and correspondingly using (B2.5),

$$[\Delta S]_{CA} = [V][Z] \begin{pmatrix} 0 & 0 & y_1 \\ 0 & 0 & y_2 \\ 0 & 0 & y_3 - 1 \end{pmatrix} [Z]^{-1}[V]^T. \tag{B2.7}$$

The following simplifications can be made towards obtaining the analytical expression for $[\Delta S]_{CA}$. Considering that $[G]_C = a_C^2[I]$ and that $[V]^T[V] = [I]$, we get $[G']_C = a_C^2[Z]^T[Z]$, so that:

$$[G']_C = a_C^2 \begin{pmatrix} \Lambda^2 + 1 & 0 & \pm\Lambda^2 \\ 0 & 1 & 0 \\ \pm\Lambda^2 & 0 & \Lambda^2 \end{pmatrix}. \tag{B2.8}$$

So, (B2.4) can be reformulated as

$$a_C^2(\Lambda^2 + 1)y_1 = G'_{13,A} - G'_{13,C}y_3,$$
$$a_C^2 y_2 = G'_{23,A} - G'_{23,C}y_3 \quad, \tag{B2.9}$$

So that the calculation of the unknown coefficients $y_1$ and $y_2$ requires 13 and 23 elements of the matrix $[G']_A - y_3[G']_C$. This matrix can be obtained as follows

$$[G']_A - y_3[G']_C = [Z]^T[V]^T[\Delta G]_{AC}[V][Z] + (1-y_3)[G']_C. \tag{B2.10}$$

The first term in (B2.10) can be calculated by considering that the columns of $[V]$ are the eigenvectors of the $[\Delta G]_{AC}$ matrix with the eigenvalues $\lambda_1, 0, \lambda_3$, so that:

$$[V]^T[\Delta G]_{AC}[V] = \begin{pmatrix} \lambda_1 & 0 & 0 \\ 0 & 0 & 0 \\ 0 & 0 & \lambda_3 \end{pmatrix},$$

and

$$[Z]^T[V]^T[\Delta G]_{AC}[V][Z] = \begin{pmatrix} 0 & 0 & \mp\lambda_1 \\ 0 & 0 & 0 \\ \mp\lambda_1 & 0 & -\lambda_1 \end{pmatrix}. \tag{B2.11}$$

For the case when AS is formed by the macroscopic averaging within the $T_a(110)T_b$ assembly (as e.g. explained in Figure 1) and $\omega_b = \omega_1 \approx \frac{x_c}{x_T}$, we use (12) for $\lambda_1$ and rewrite (B2.11) as

$$[Z]^T[V]^T[\Delta G]_{AC}[V][Z] \approx 2a_C^2 \begin{pmatrix} 0 & 0 & \mp x_a \\ 0 & 0 & 0 \\ \mp x_a & 0 & -x_a \end{pmatrix}. \tag{B2.12}$$

The second term of (B2.10) can be derived using expression (20) for $y_3$ and (B2.8), so that

$$(1-y_3)[G']_C \approx -(2x_a + x_c)a_C^2 \begin{pmatrix} M^2 + 1 & 0 & \pm M^2 \\ 0 & 1 & 0 \\ \pm M^2 & 0 & M^2 \end{pmatrix}. \tag{B2.13}$$

According to (17), $M^2 \approx \frac{\bar{x}_a}{x_a + x_c}$, and therefore

$$\frac{(2x_a + (2x_a + x_c)M^2)}{(M^2 + 1)} = x_a. \tag{B2.14}$$

Thus, we can re-write (B2.4) as

$$\begin{matrix} y_1 \approx \mp x_a \\ y_2 = 0 \end{matrix}. \tag{B2.15}$$

Here the signs − and + correspond to the case of HP normal to $\boldsymbol{p}_1$ and $\boldsymbol{p}_2$, respectively. This result is used in the main text to obtain the expressions for the orientation relationship matrices.

**Acknowledgements:**

We acknowledge Dr Alexei Bosak for his assistance during the synchrotron experiment at the ID28 beamline at ESRF. SG acknowledges support from Israel Science Foundation (grant Nos. 1561/18, 1365/23) and United States – Israel Binational Science Foundation (award No. 2018161). SG and NZ acknowledge support of the joint Israel Science Foundation and National Science Foundation of China (grant No. 3455/21).